\documentclass[aps,pra,twocolumn,showpacs,amsmath,amssymb]{revtex4-1}
\usepackage{graphicx}

\newcommand{\ket}[1]{ {\vert #1 \rangle}}
\newcommand{\bra}[1]{ {\langle #1 \vert}}
\newcommand{\braket}[2]{ {\langle #1 \vert #2 \rangle}}

\newcommand{\xket}[1]{ {\vert #1 )}}
\newcommand{\xbra}[1]{ {( #1 \vert}}
\newcommand{\xbraket}[2]{ {( #1 \vert #2 )}}

\begin{document}

\author{Iztok Pi\v{z}orn}
\affiliation{Theoretische Physik, ETH Zurich, CH-8093 Z\"urich, Switzerland}

\date{August 14, 2013}

\pacs{
21.60.Fw, % Bosons interacting boson model, 21.60.Fw
05.30.Jp, % Boson systems, 05.30.Jp
05.70.Ln, % Nonequilibrium processes  in thermodynamics
03.75.Gg, % Decoherence : Bose-Einstein condensates
42.50.Dv, % Photons nonclassical states, 42.50.Dv
32.80.Xx, % Optical pumping of atoms
42.50.Pq % Cavity QED 
}

\title{Bose Hubbard model far from equilibrium}

\begin{abstract}
We study the nonequilibrium steady state of the Bose Hubbard model coupled to Lindblad reservoirs,  using the density matrix renormalization group in operator space. We observe a transition from a flat particle density profile in the noninteracting limit to the linear profile with onset of the interparticle interaction. Analyzing the effect of coherent pumping on the nonequilibrium steady state we find a subspace which remains unaffected by the pumping in the noninteracting limit with the protection gradually diminishing due to interparticle interaction. 
In the equilibrium situation with one or more symmetric reservoirs we show analytically that the steady state of the system is a product state for any interaction strength. We also provide analytical results in the noninteracting limits, using the method of the third quantization in operator space. 
\end{abstract}

\maketitle

\section{Introduction}
Synergy of physics of quantum fluids, quantum information and quantum computing has introduced a new paradigm in studying properties of quantum many-body systems. 
Essentially limitless capabilities in manipulation of optical lattices \cite{vahala,blochnature,blochreview}
which can address atoms with a single site resolution by optical imaging \cite{bakr,sherson} or scanning electron microscopy \cite{gericke,wurtz}, allow precise tuning of control parameters. 
The discovery that the celebrated Bose Hubbard model
can be realized by cold-atom optical lattices \cite{jaksch}, and its experimental confirmation \cite{greiner}, combined with great control capabilities in optical lattices, makes cold atoms an ideal setting for exploring properties of strongly correlated condensed matter systems \cite{lewenstein}, in particular at equilibrium due to remarkably good isolation from the environment. Manipulation of the system, however, results in heating due to incoherent scattering of laser light \cite{pichler}, or particle loss due to interaction of atoms  with the focussed electronic beam. The incurred particle loss can have interesting consequences and can in fact pronounce the correlations \cite{syassen,kiffner}, generate entanglement \cite{barmettler}, enhance the coherence of the system \cite{trimborn,witthaut,kordas} and result in nonequilibrium quantum phase transitions \cite{diehlprl}.

Similarly to cold-atom optical lattices, the condensed matter systems can be realized also in 
exciton polariton gases \cite{carusottociuti2004} and photon gases \cite{greentree,hartmannplenio,angelakis,hartmann,carusotto,ciutireview}, 
implemented as arrays of nonlinear cavity resonators.
Unlike with cold-atoms, dissipation processes are unavoidable in photonic systems and photon loss must be compensated by pumping \cite{ciutireview}. Photonic lattices are thus intrinsically nonequilibrium systems and offer a promising setting to explore \emph{open} quantum many-body systems. 
The dynamics of coupled nonlinear cavities can be described by a dissipative Bose Hubbard-like model incorporating coherent pumping, known as 
 \textit{driven-dissipative} Bose Hubbard model \cite{drummond,ferretti,jin,leboite} where the dynamics is determined by the interplay of coherent laser pumping and dissipation and leads to various phases of the steady state \cite{jin,leboite,leib}.

All these features of nonequilibrum steady states shift the role of dissipation from the unavoidable and undesired phenomenon where dissipation is overcome with help of decoherence-free subspaces \cite{zanardi,lidar}, to a useful resource which can be exploited in universal quantum computation \cite{verstraeteopen,diehl,kraus,kastoryano,diehlrico} by controlling both dissipation and Hamiltonian dynamics \cite{barontini}. 

Coupling the system to the environment does not only result in particle losses, but also enables injection of particles, e.g. by incoherent pumping in optical lattices of photons \cite{ciutireview,szymanska}. 
Open one-dimensional systems where sites on either end are coupled to reservoirs have been subject of several studies, focusing in particular to transport properties. The dynamics in the Markovian approximation for the reservoirs is given by the Lindblad master equation which can be simulated numerically using the time dependent Density Matrix Renormalization Group algorithms (t-DMRG) \cite{schollwoeck,daley} in operator space \cite{verstraetempo,zwolak,prosenznidaricness}. 
This matrix product state representation has been used to study transport properties of quantum  chains \cite{benenti,marko2,marko3,marko4,prosenznidaricqpt,mendozax,mendoza}, and also resulted in exact results in some cases, see e.g. Refs. \onlinecite{prosenxxz1,prosenxxz2,karevski}.
Furthermore, an explicit solution for Liouville dynamics of quadratic quantum many-body systems with linear Lindblad reservoirs was proposed \cite{prosen3q} (so called ``third quantization'') and recently extended to bosonic systems \cite{prosenseligman3q}. 

In light of these results for nonequilibrium effective spin systems, we analyze the steady state of the Bose Hubbard model with Lindblad reservoirs attached to either end of the system. Unlike other approaches to nonequilibrium Bose Hubbard-like models, we consider both dissipation and pumping of particles, resulting in a nontrivial steady state.
In the noninteracting limit and also in the setting with one or many symmetric reservoirs we provide analytic solutions to the particle density profile by means of the method of third quantization and operator space formalism, respectively, whereas the generic asymmetric case is simulated by means of the tDMRG algorithm. 
We also study the steady state of the driven-dissipative Bose Hubbard model and observe a subspace protected to coherent pumping in the noninteracting limit which gradually disappears with onset of interaction.

\section{Model}
The Bose Hubbard model on a one-dimensional lattice of $n$ sites 
is described by a Hamiltonian operator 
\begin{equation}
H_{\rm BH} = -t \sum_j (a_j^\dagger a_{j+1} + a_{j+1}^\dagger a_j) + \frac{U}{2} \sum_j n_j (n_j-1) 
\label{eq:H}
\end{equation}
where the tunneling term at rate $t$ corresponds to the kinetic energy and $U$ is the strength of the onsite repulsive interaction between particles.
The annihilation and creation operators,  denoted as $a_j$ and $a_j^\dagger$, respectively, satisfy $[a_j, a_l^\dagger] = \delta_{jl}$ and $[a_j,a_l]=0$ whereas $n_j = a_j^\dagger a_j$ denotes the local particle number operator. The Bose Hubbard model can be realized in cold-atom optical lattices \cite{blochreview} as well as in photon gases \cite{ciutireview} where both parameters can be precisely controlled. 
We shall also consider an additional Hermitian term corresponding to coherent pumping in photon gases \cite{ciutireview}
\begin{equation}
H = H_{\rm BH} + \Omega \sum_j (a_j^\dagger + a_j)
\label{eq:HOmega}
\end{equation}
where $\Omega$ is the coherent pumping amplitude. 
Unlike the standard Bose Hubbard model~(\ref{eq:H}), the extended model~(\ref{eq:HOmega}) does not preserve the number of particles in the system.

Let us now introduce dissipation by coupling the system to a reservoir. After tracing out the degrees of freedom in the environment and assuming Markovian reservoirs, the effect of the coupling to the reservoirs is described by a master equation in the Lindblad form 
\begin{equation}
(d/dt) \rho = -i [H, \rho] + 
\sum_{\mu} \big( 2 L_\mu \rho L_{\mu}^\dagger - \{ L_{\mu}^\dagger L_{\mu},\rho \} \big)
\equiv \mathcal{L}(\rho)
\label{eq:Lrho}
\end{equation}
where $L_{\mu}$ are Lindblad operators.
\begin{figure}
\includegraphics[width=\columnwidth]{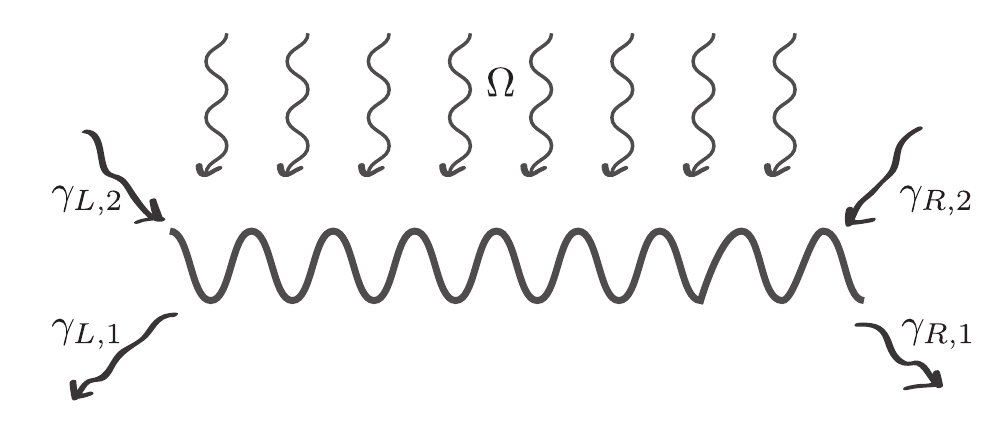}
\caption{Nonequilibrium Bose Hubbard model coupled to Markovian reservoirs on either end, representing dissipation (with rate $\gamma_{L,1}$ and $\gamma_{R,1}$ for left and right, respectively) and incoherent pumping (with rate $\gamma_{L,2}$ and $\gamma_{R,2}$) of particles, as well as coherent homogeneous pumping with an amplitude $\Omega$.}
\label{fig:sketch}
\end{figure}
We shall assume that the system can exchange particles with the environment in which case the Lindblad operators are given as 
\[
L_{(1,j)} = \sqrt{\gamma_{1,j}}  \, a_j
\textrm{ and }
L_{(2,j)} =\sqrt{\gamma_{2,j}}  \, a_j^\dagger
\]
where $L_{(1,j)}$ is associated with emission of a particle at the $j$-th site in the system with rate $\gamma_{1,j}$ whereas $L_{(2,j)}$ corresponds to absorption. 
The dynamics of a mixed state system~(\ref{eq:Lrho}) is thus described by the unitary (dissipation free) evolution $-i [H, \rho]$ and the non-unitary Lindblad dissipative part.

In the course of this work we shall assume that after a long time the system evolves towards a nonequilibrium steady state $\rho_{\rm ness}$ defined as a solution to $(d/dt) \rho \vert_{\rho=\rho_{\rm ness}} = 0$  or equivalently 
\begin{equation}
\mathcal{L}(\rho_{\rm ness}) = 0.
\label{eq:steadystate}
\end{equation}
The existence of a steady state is not guaranteed in bosonic systems. In case where the emission rate is higher than the absorption rate, 
$\gamma_{2,j} > \gamma_{1,j}$, the system could accumulate more and more particles and no steady state would exist. 
We shall therefore restrict our study to the case $0 \geq \gamma_{2,j}/ \gamma_{1,j} < 1$. 
Similarly, when the system is purely dissipative and the loss of particles is not compensated by 
pumping ($\gamma_{2,j}=0$ of incoherent and $\Omega=0$ for coherent pumping), 
the steady state is trivial and is equal to the vacuum pure state, except in the case of decoherence-free subspaces.

\section{Method}
In general, the determination of the steady state~(\ref{eq:steadystate}) is a quantum many-body problem where the computational complexity scales exponentially with the system size. Nevertheless, the Liouvillian operator $\mathcal{L}$ defined in ~(\ref{eq:Lrho}) is a linear operator and the question of the steady state can be reduced to an infinite dimensional linear algebra problem. 
In the noninteracting limit $U=0$, the steady state can be constructed explicitly using the formalism of the ``third quantization'' \cite{prosen3q,prosenseligman3q} and the expectation value of any local observable can be calculated exactly. 
In the generic interacting case, however, we must resort to numerical simulations.
For these we shall assume a truncated $K$-dimensional bosonic space by allowing at most $K-1$ particles per site. We shall work in operator space such that the master equation with a Liouville operator $\hat{\mathcal{L}}$ is expressed in a form resembling the Schr\"odinger equation,
\begin{equation}
(d/dt) \xket{\rho} = \hat{ \mathcal{L} } \xket{\rho}.
\label{eq:rhot}
\end{equation}
This in turn allows us to simulate the steady state by the time evolution in operator space
\[
\xket{\rho_{\rm ness} } = \lim_{t\to\infty} e^{t\hat{\mathcal{L}} } \xket{\rho_0}
\]
using the time-dependent DMRG algorithm \cite{prosenznidaricness}. 
If the steady state, which is a right null vector of the Liouvillian operator $\hat{\mathcal{L}}$ such that $\hat{\mathcal{L}} \xket{\rho_{\rm ness}} = 0$,  is unique, any initial state $\rho_0$ will after a sufficiently long time relax to the steady state $\rho_{\rm ness}$, as long as it has a nonzero overlap with the \emph{left} null vector which is always equal to the identity operator, $\xbra{\mathbf{1}} \hat{\mathcal{L}} =0$, 
i.e. $\xbraket{\mathbf{1}}{\rho_0} \neq 0$. In particular, we start the simulation with the infinite-temperature equilibrium state $\rho_0 = \mathbf{1}$ and evolve it corresponding to~(\ref{eq:rhot}) ad libitum.

The density matrices are represented as matrix product states in operator space
\begin{equation}
\xket{\rho} = \sum_{\underline{\alpha}} {\rm tr}\Big( \prod_{j=1}^{n} \mathbf{A}^{[j] \alpha_j}  \Big)
\xket{ g^{\alpha_1} \otimes \cdots \otimes g^{\alpha_n} }
\label{eq:mps}
\end{equation}
where $\{ g^{\nu}; \enskip \nu=1,2,\ldots,K^2 \}$ are generalized $K\times K$ Gell-Mann matrices  and $K$ is the dimension of the local (truncated) bosonic space (the dimension of operator space is $K^{2n}$). Throughout the paper we shall adopt a vector notation $\underline{a} = (a_1,a_2,\ldots)$.
Due to the hermiticity and orthogonality of basis states (direct products of Gell-Mann matrices),
we can choose $\mathbf{A}^{[j] \alpha_j}$ in~(\ref{eq:mps}) as real matrices which reduces the accessible operator space to Hermitian (but not neccessarily positive definite) operators.

Practical implementations of the algorithm require a drastic truncation of the bosonic space. While the lowest nontrivial choice $K=3$ already provides a good qualitative picture, a faithful representation requires a slightly larger local bosonic space with $K = 6 \sim 8$. 
The complexity of numerical simulation is dominated by singular value decompositions of dense $(K^2 D) \times (K^2 D)$ matrices where however only the leading $D$ singular vectors are retained. This  stimulates the usage of iterative SVD schemes, such as the \emph{randomized SVD} \cite{redsvd} or the Lanczos algorithm which both scale as $O(K^4 D^3)$ as opposed to $O(K^6 D^3)$ in the full (e.g. QR-based) SVD implementation. We shall typically deal with low correlated states which are well described already by $D\sim 32$ which allows us to work with $K \sim 7$. An accurate time evolution of the fully mixed density operator towards the steady state would in principle require larger bond dimensions in the transient regime. We are however only interested in the final steady state in which case working in a low correlated subspace of operator space yields the same results.

Our principal interest lies in the particle density profile $\langle n_j  \rangle$ of the steady state 
which (in operator space framework) is given as a scalar product with the density operator, 
\begin{equation}
\langle n_j \rangle = \xbraket{n_j }{\rho_{\rm ness}} / \xbraket{\mathbf{1}}{\rho_{\rm ness}}.
\end{equation}
Other quantities can be calculated in a similar way, $\langle a \rangle = \xbraket{a}{\rho_{\rm ness}} / \xbraket{\mathbf{1}}{\rho_{\rm ness}}$.

Another quantity of interest is the amount of correlations in the steady state which we will quantify
through the operator space entanglement entropy (OSEE) \cite{prosenpizornpra}, 
defined as an operator space analogue of the  entanglement entropy for quantum states as
\begin{equation}
S^\sharp = -{\rm tr}( \hat{R} \log \hat{R} ) 
\label{eq:osee}
\end{equation}
where $\hat{R}$ is the ``reduced density operator'' in operator space, obtained by summing over the degrees of freedom in one part of the system, 
\[
\hat{R} = {\rm tr}_{n/2+1,\ldots,n} \xket{\rho_{\rm ness}}\xbra{\rho_{\rm ness}}.
\]

The OSEE is obtained from a MPS description of the density operator in a straight forward way from 
Schmidt coefficients $\underline{\lambda}$ in a bipartite splitting of the MPS as
$S^\sharp = - \underline{\lambda} \cdot \log(\underline{\lambda})$, 
assuming $\underline{\lambda} \cdot \underline{\lambda} = 1$.
It was shown that the OSEE can be used as a signature of quantum critical behavior of the nonequilbrium steady state, being infinite in the thermodynamic limit in the case of steady states with long range correlations  and finite otherwise \cite{prosenpizornprl}.

\subsection{Liouvillian operator in operator space notation}
The Liouvillian operator is a map over operator space which itself is infinitely dimensional also for a finite lattice size. In order to give a concise mathematical description, we shall adopt the notation used in Refs.~\onlinecite{prosen3q,prosenseligman3q} where the operator space is presented in a form of a Fock space, generated by operators $\hat{a}_{\nu,j}$ and $\hat{a}_{\nu,j}'$ for $\nu \in \{0,1\}$ which satisfy almost canonical relations 
\[
[\hat{a}_{\mu,j}, \hat{a}_{\nu,l}'] = \delta_{\mu,\nu} \delta_{j,l}
\quad\textrm{and}\quad
[\hat{a}_{\mu,j}, \hat{a}_{\nu,l}] = [\hat{a}_{\mu,j}', \hat{a}_{\nu,l}'] = 0
\]
and $\hat{a}_{\mu,j}' \neq \hat{a}_{\nu,j}^\dagger$.
In the course of this paper we shall use the hat notation $\hat{a}$ exclusively when referring to maps \emph{over} operator spaces whereas maps over the Hilbert space (the operators) will be written without a hat.
For bosonic systems, one way to choose $\underline{\hat{a}}$ and $\underline{\hat{a}}'$ is 
\begin{equation}
\begin{matrix}
\hat{a}_{0,j} = \hat{a}_j^{\rm L} & \quad &
\hat{a}_{0,j}' = \hat{a^\dagger}_j^{\rm L} - \hat{a^\dagger}_j^{\rm R} 
\\
\hat{a}_{1,j} = \hat{a^\dagger}_j^{\rm R} & \quad &
\hat{a}_{1,j}' = \hat{a}_j^{\rm R} - \hat{a}_j^{\rm L}
\end{matrix}
\label{eq:anuj}
\end{equation}
where $\hat{b}^{L}$ and $\hat{b}^R$ map an element $y$ of operator space to
$b y$ and $y b$, respectively. For example, $\hat{a_j^\dagger}^R \xket{y} = \xket{y a_j^\dagger}$. 
In analogy to the hat-convention, 
we use the braket notation $\xket{y}$ when referring to operators as elements of operator space.

In this notation, the density matrix $\rho$ is written as a superposition of Fock (operator-)states 
\[
\xket{\rho} = \sum_{\underline{m} } \rho_{\underline{m}} \Big( \prod_{j=1}^{n} \frac{  (\hat{a}_{0,j}')^{m_{0,j}} (\hat{a}_{1,j}')^{m_{1,j}} }{\sqrt{ m_{0,j}! m_{1,j}! } } \Big)
\xket{\rho_0}
\]
where $\rho_0$ is the density matrix describing the vacuum pure state 
$\ket{0}\bra{0}$ for which $a_j \ket{0} =0$ and thus $\hat{a}_{\nu,j} \xket{\rho_0} = 0$. 
The Liouvillian operator~(\ref{eq:Lrho}) can now be written as a non-quadratic ``super-operator'' in terms of maps~(\ref{eq:anuj}) as 
\begin{eqnarray}
\hat{\mathcal{L}} &=& 
   i t \sum_j (\hat{a}_{0,j}' \hat{a}_{0,j+1} 
                 + \hat{a}_{0,j+1}' \hat{a}_{0,j}  
                  - \hat{a}_{1,j+1}' \hat{a}_{1,j}
                  - \hat{a}_{1,j}' \hat{a}_{1,j+1}  ) 
                   \nonumber \\
&-& i \frac{U}{2}\sum_j (\hat{n}_{0,j} + \hat{n}_{1,j} 
                             + 2 \hat{a}_{0,j} \hat{a}_{1,j} -1) 
                             ( \hat{n}_{0,j} - \hat{n}_{1,j}) \nonumber\\
&-& i \Omega \sum_j (\hat{a}_{0,j}' - \hat{a}_{1,j}) \\
&+&  \sum_j 
[ (\gamma_{j,1} - \gamma_{j,2}) (\hat{n}_{0,j}+\hat{n}_{1,j}) -
                        2 \gamma_{j,2} \hat{a}_{0,j}' \hat{a}'_{1,j} ]. \nonumber
\label{eq:liouvillian}
\end{eqnarray}
The first three rows in the above expression form the unitary part of the evolution whereas the last row represents dissipative part in the Lindblad form.
We use an abbreviation $\hat{n}_{\nu,j} = \hat{a}_{\nu,j}' \hat{a}_{\nu,j}$ which multiples $(\hat{a}_{\nu,j}')^{m} \ket{\rho_0}$ by a factor of $m$ and thus plays the role of the number operator in operator space.

%%%%%%%%%%%%%%%%%%%%%%%%%%%%%%%%%%%%
% RESULTS
%%%%%%%%%%%%%%%%%%%%%%%%%%%%%%%%%%%%

\section{Results}
We investigate the nonequilibrium steady state of the one dimensional (driven-dissipative) Bose Hubbard model in a presence of reservoirs on either end of the system.
We consider a one-dimensional lattice of $n=16$ sites with a bosonic truncation parameter $K = 6,7$. 
Full description of a density matrix would require of the order of $K^{2n}$ parameters, which we 
parametrize by a real matrix product state with a bond dimension $D=100$. 
Typically, the bond dimension of the final steady state is even lower ($D \sim 50$); intermediate dynamics of the Lindblad master equation requires more accurate description.

\subsection{Steady state of dissipative Bose Hubbard model}
We start by studying the Bose Hubbard model in dissipative regime without coherent pumping ($\Omega=0$). In this regime, the steady state is trivial unless we allow for the absorption of particles by imposing $\gamma_{2,L} > 0$ and (or) $\gamma_{2,R} > 0$.

\subsubsection{Noninteracting limit}
\begin{figure}
\includegraphics[width=\columnwidth]{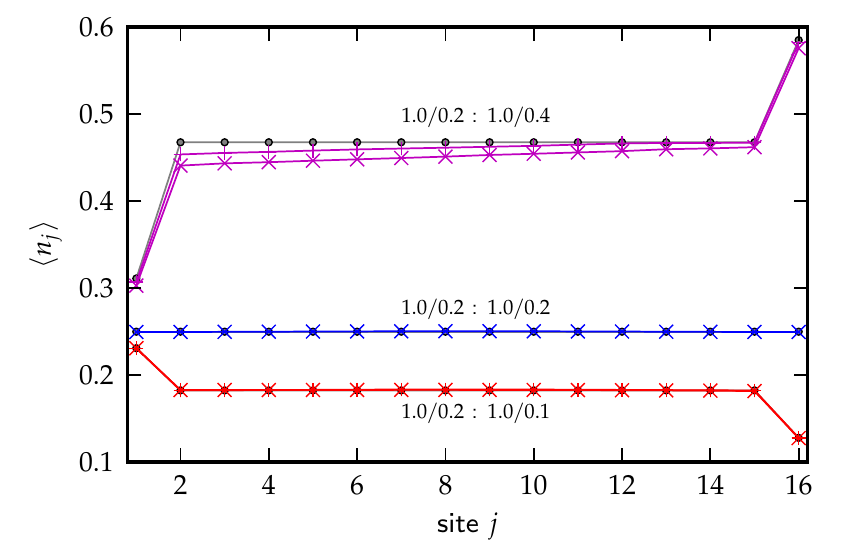}
\caption{(Color online) Particle density profile of the steady state for the noninteracting bosonic model ($U=\Omega=0$, $t=0.5$) with a fixed left reservoir $\gamma_{L} = (1,0.2)$ and various right reservoirs $\gamma_R = (1,0.1), (1,0.2), (1,0.4)$ as denoted in the plot. Numerical results for $K=6$ (cross) and $K=7$ (plus) are compared to exact results shown in grey (circle). }
\label{fig:bhex4A}
\end{figure}

The noninteracting case $U=0$ allows for an exact solution which well agrees with our numerical results obtained by the time-dependent DMRG in operator space as seen from Fig.~\ref{fig:bhex4A} where we show the particle density profile for a fixed left reservoir and three different right reservoirs. 
When the reservoirs are symmetric, meaning that $\underline{\gamma}_L = \underline{\gamma}_R$ (it actually sufficies $\gamma_{2,L}/\gamma_{1,L} = \gamma_{2,R}/\gamma_{1,R}$ as we shall see later), the density profile is flat.  Due to the integrability, it is not surprising that the density profile remains flat in the bulk also in the nonequilibrium steady state with different reservoirs on each end of the system. Numerical data can be compared to exact results obtained in the framework of the third quantization as shown in Fig.~\ref{fig:bhex4A} where we observe that bosonic truncation parameters $K=6,7$ (colored lines) 
give the same qualitative picture as the exact results (plotted in grey).

The exact solution for the noninteracting case can be obtained analytically. For convenience, we shall introduce a new set of reservoir parameters
\begin{equation}
\kappa_j \equiv \gamma_{j,1}/\gamma_{j,2} - 1
\quad\textrm{and}\quad
\Gamma_j \equiv \gamma_{j,2} / t
\label{eq:kappa}
\end{equation}
where $\kappa_j$ measures the ratio between the emission and the absorption rate whereas $\Gamma_j$ measures the strength of couplings to the reservoirs with respect to the tunneling rate.
The particle density in the bulk can be calculated after a lengthy but straight forward calculation (see Appendix \ref{app:A}) and reads
\[
\langle n_j \rangle_{\rm bulk} = 
\frac{
\Gamma_L \Gamma_R (\Gamma_L \kappa_L^2 + \Gamma_R \kappa_R^2) + 
(\Gamma_L + \Gamma_R)
}{
(\Gamma_L \kappa_L + \Gamma_R \kappa_R) (\Gamma_L \Gamma_R \kappa_L \kappa_R + 1)
  }.
\]
The particle density is different at the boundary sites which are coupled to reservoirs and for the left site it is given as
\[
\langle n_1 \rangle = 
\frac{ \kappa_R \Gamma_L \Gamma_R  (\Gamma_L \kappa_L  + \Gamma_R \kappa_R) 
+ (\Gamma_L + \Gamma_R) 
}{(\Gamma_L \kappa_L +  \Gamma_R \kappa_R) (\Gamma_L \Gamma_R \kappa_L \kappa_R + 1)
  }
\]
and approaches the value $\kappa_L^{-1}$ in the limit of weak tunneling rate $t \to 0$. 
A similar result is obtained for the right-most site by exchanging indices $L$ and $R$ in the above expression.

In the opposite limit of weak coupling to the reservoirs, 
the density profile approaches the equilbrium value resembling a weighted average of reservoir-imposed expectation values at the boundaries, 
\begin{equation}
\lim_{t \to\infty} \langle n_j \rangle \to 
\Big( 
\frac{\Gamma_L}{\Gamma_L+\Gamma_R} \kappa_L + 
\frac{\Gamma_R}{\Gamma_L+\Gamma_R} \kappa_R
\Big)^{-1}.
\label{eq:weakcoupling}
\end{equation}
The particle density profile is in this limit of strong tunneling constant throughout the system, including at the boundary sites which are coupled to reservoirs.

Here we make a connection to a related study of a quantum chaotic coupled to two thermal reservoirs with different temperatures on either end \cite{weidenmueller} where it was shown that the steady state for equal couplings to the reservoirs (related to $\Gamma_L = \Gamma_R$ in our case) leads to a thermal steady state of the system with an intermediate temperature being equal to the mean temperature of both reservoirs and the weighted mean temperature in the case of similar couplings to the reservoirs. This is reminiscent to our result which is however obtained for an \emph{integrable} model in the weak coupling limit~(\ref{eq:weakcoupling}).

\subsubsection{Interacting Bose Hubbard model}
\begin{figure}
\includegraphics[width=\columnwidth]{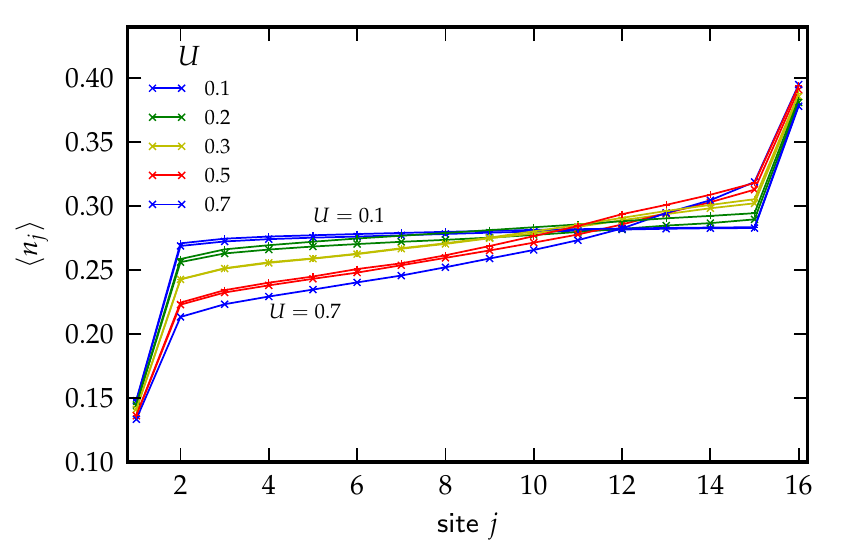}
\caption{(Color online) Particle density profile of the steady state for the Bose Hubbard model on $n=16$ sites ($t=0.5$, $\Omega=0$) with various interaction strengths $U=0.1,\ldots,0.7$ (from top to bottom on the left) and asymmetric reservoirs on either edge with rates $\gamma_L=(1,0.1)$ and $\gamma_R=(1,0.3)$. Bosonic truncation parameter used was $K=6$ (cross) and $K=7$ (plus). }
\label{fig:bhex4B}
\end{figure}
Let us now consider the interacting Bose Hubbard model coupled to reservoirs on the edges. 
The equilibrium situation with a single reservoir or many symmetric reservoirs with the same ratios $\{ \kappa_j   = \kappa\}$ results in a steady state which can be written as a product state, regardless of the interaction strength $U$. A straight forward calculation (see Appendix~\ref{app:B}) yields a particle density 
\[
\langle n_j \rangle = \kappa^{-1}\quad j=1,2,\ldots,n.
\]
Such a state is completely uncorrelated and corresponds to the infinite temperature equilibrium state for a grand-canonical ensemble \cite{markoprivate}
$e^{- \beta (H - \mu N)}$ where
\[
\beta \to 0 
\quad\textrm{and}\quad
\beta \mu = \kappa^{-1}.
\]
This surprising result tells us that there always exists a noncorrelated steady state of the Bose Hubbard model with a one reservoir towards which the system will evolve from any initial starting state, provided that the steady state is unique. The uniqueness of the steady state is however not guranteed in general and there might exist a whole manifold of steady states, forming the decoherence free subspace. This is indeed the case in the noninteracting limit with an \emph{odd} number of sites and a single reservoir attached to any \emph{even} site where an infinitely dimensional manifold of steady states exists. This is a generalization and refinement of the result in Ref.~\onlinecite{kepesidis} where a decoherence free subspace was found by considering systems with an odd number of sites and a dissipative reservoir in the center of the system. 

\begin{figure}
\includegraphics[width=\columnwidth]{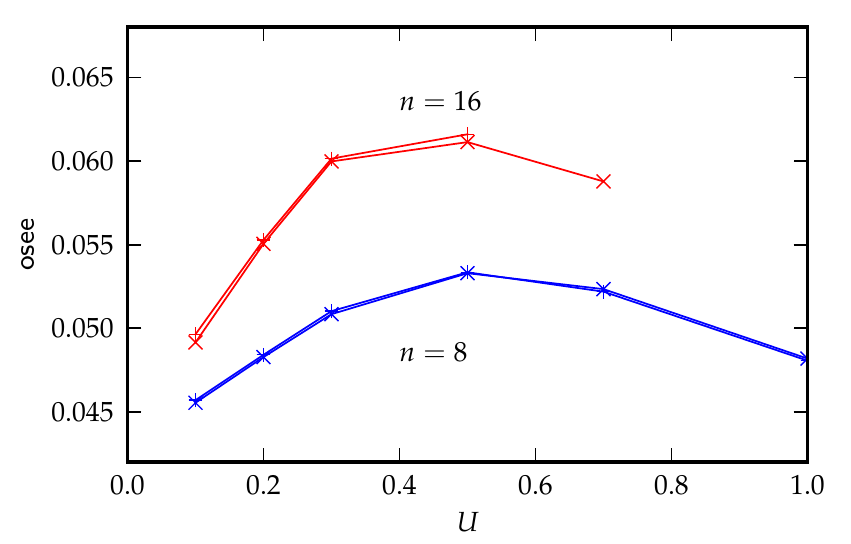}
\caption{(Color online) Operator space entanglement entropy of the NESS for dissipative Bose-Hubbard model on $n=8$ (blue) and $n=16$ (red) sites with fixed reservoirs $(1,0.1)$ and $(1,0.3)$ without coherent pumping $\Omega=0$, as a function of interaction strength $U$. Bosonic truncation parameters used were $K=6$ (cross) and $K=7$ (plus).}
\label{fig:bhex4X}
\end{figure}
Let us now focus on the nonequilibrium situation with two different reservoirs on either edge of the system. We choose the reservoir parameters as 
$\underline{\gamma}_L = (1,0.1)$ and $\underline{\gamma}_R = (1,0.3)$ which translates to
$\kappa_L^{-1}  \approx 0.11$ and $\kappa_R^{-1} \approx 0.43$; any other choice $\kappa_L \neq \kappa_R$ would result in a similar qualitative picture.
In Fig.~\ref{fig:bhex4X} we show the particle density profile for various values of the interaction strength $U$ and the system size $n=16$. The bosonic degrees of freedom were truncated to $K=7$. 
We observe that the initially flat particle density profile in the bulk for the noninteracting case gradually evolves to a linear profile in the bulk with onset of interparticle interaction, and eventually tends to an overall linear profile between values
$\langle n_1 \rangle = \kappa_L^{-1}$ on the left and $\langle n_n \rangle = \kappa_R^{-1}$ on the right, as imposed by the reservoirs. 

We also study the degree of correlations in the system as measured by OSEE and observe that the system is most correlated a some intermediate value of the interaction strength as shown in Fig.~\ref{fig:bhex4X}. The result is independent of the bosonic truncation parameters and the results for $K=6$ and $K=7$ overlap.

\subsection{Steady state of driven-dissipative Bose Hubbard model}
In addition to incoherent pumping, particle loss in the system can be compensated also by coherent pumping as is the case in photon gases, represented by an additional Hermitian term in the Hamiltonian operator, given in~(\ref{eq:HOmega}). 
The parameter $\Omega$ denotes the coherent pumping amplitude.
In this case, the total particle number is not a conserved quantity even in Hamiltonian dynamics which leads to interplay between dissipation and incoherent pumping on one side and coherent pumping on the other.

\subsubsection{Noninteracting regime}
\begin{figure}
\includegraphics[width=\columnwidth]{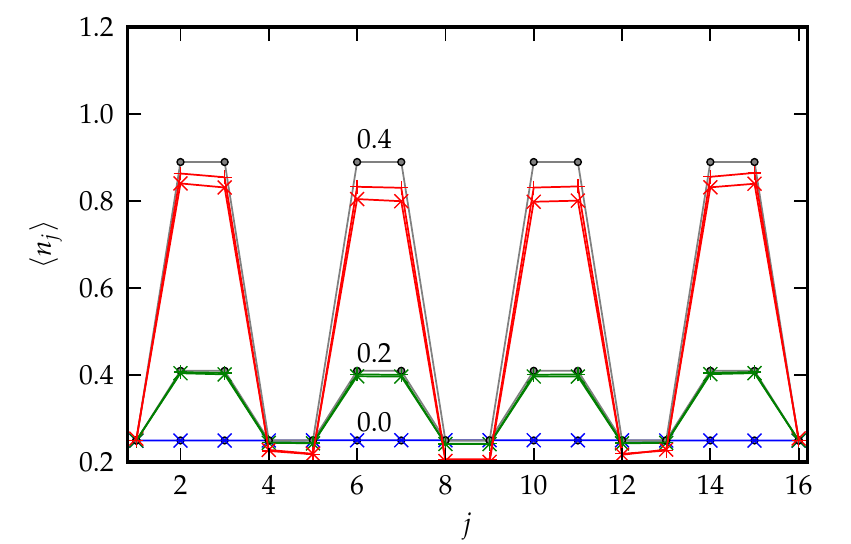}
\includegraphics[width=\columnwidth]{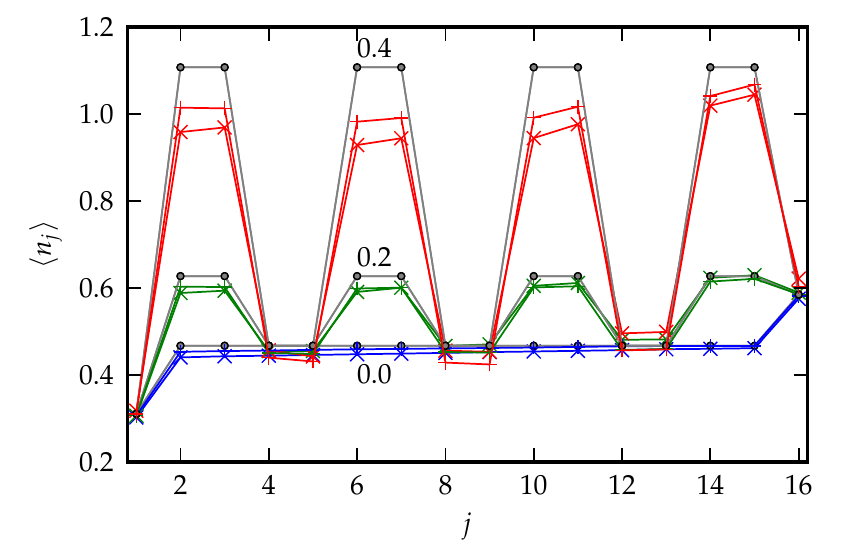}
\caption{(Color online) Particle density profile of the NESS for a noninteracting driven-dissipative model on $n=16$ sites ($U=0$, $t=0.5$) and various coherent pumping amplitudes $\Omega=0,0.2,0.4$, as annotated in the plots.
The reservoirs on either end are taken symmetrically (top) with $\underline{\gamma}_L = \underline{\gamma}_R = (1,0.2)$,  or asymmetrically (bottom) with $\underline{\gamma}_L = (1,0.2)$ and $\underline{\gamma}_R = (1,0.4)$.
Bosonic truncation parameters used were $K=6$ (cross) and $K=7$ (plus).
The exact results are shown in grey (circle). }
\label{fig:bhex4C}
\end{figure}
We shall again start with the noninteracting case which we can solve analytically by means of the third quantization (see Appendix~\ref{app:A}). 
We calculate the particle density profile in the steady state for $n=16$ sites and various pumping amplitudes $\Omega$. We consider both symmetric  and asymmetric reservoirs on either end of the system and observe a distinct pattern behavior where a sublattice of the system is protected from  coherent pumping, as shown in Fig.~\ref{fig:bhex4C}.
The effect of coherent pumping is visible as a constant addition (see Appendix~\ref{app:A})
\[
\langle n_j \rangle = \langle n_j \rangle\big\vert_{\Omega=0} + 
\begin{cases} 
\Omega^2 / t^2 & \text{if $j=2,3,6,7,10,11,\ldots$,} \\
0 & \text{if $j=1,4,5,8,9,\ldots$.}
\end{cases}
\]
to the density profile, on top of the solution for the pumping-free nonequilibrium steady state,  exhibiting a flat density profile in the bulk (cf. Fig.~\ref{fig:bhex4B}). The exact analytic result is plotted in grey and numerical simulations require keeping relatively many bosonic degrees of freedom (i.e. more than $K=7$ considered in our simulations), despite a relatively low expectation value of the particle density of the order of one particle per site.

The above result is however only correct when the number of sites is divisible by four. Otherwise, the protection is not strict but the pattern shown in Fig.~\ref{fig:bhex4C} persists (see Appendix~\ref{app:A}). 
This phenomenon is similar to the decoherence-free subspace discussed previously in the case of one reservoir coupled to the Bose Hubbard chain, however, here the system is protected from coherent pumping instead of dissipation.

\subsubsection{Interacting regime}
\begin{figure}
\includegraphics[width=\columnwidth]{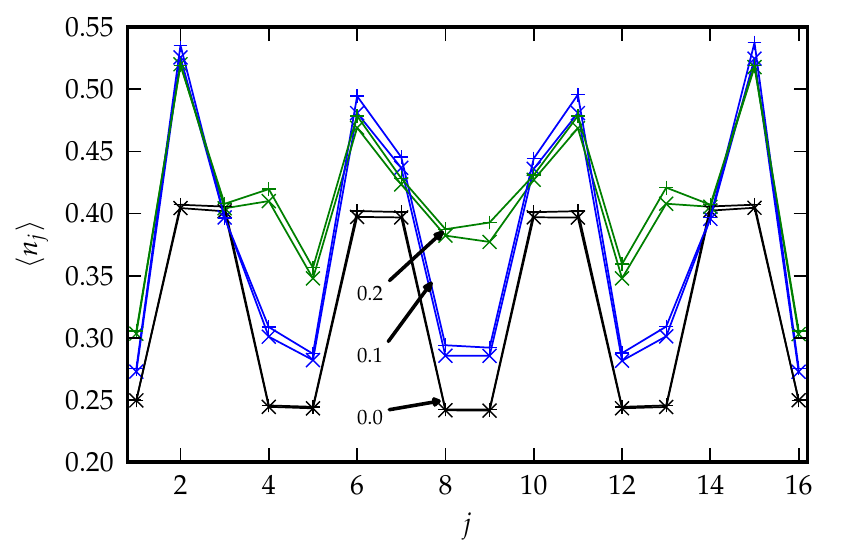}
\caption{(Color online) Particle density profile of the NESS for a driven-dissipative Bose Hubbard model on $n=16$ sites ($t=0.5$) with a fixed coherent pumping amplitude $\Omega=0.2$ and interaction strengths $U=0,0.1,0.2$, annotated in the plot. The reservoirs are symmetric on either edgee, $\underline{\gamma}_L = \underline{\gamma}_R=(1,0.2)$.
Bosonic truncation parameters used were $K=6$ (cross) and $K=7$ (plus).
}
\label{fig:bhex4F}
\end{figure}

In the noninteracting case we calculate the steady state properties numerically by means of t-DMRG simulations. 
In Fig.~\ref{fig:bhex4F} we show the particle density profiles for a fixed coherent pumping amplitude $\Omega$ and various strengths of interparticle interaction $U$.
We observe that the subspace protection to coherent pumping disappears with onset of interparticle interaction and gradually approaches a smooth density profile.

%%%%%%%%%%%%%%%%%%%%%%%%%%%%%%%%%%%%%%%%%%%%%
% Conclusions
%%%%%%%%%%%%%%%%%%%%%%%%%%%%%%%%%%%%%%%%%%%%%

\section{Conclusion}
We have investigated the noneqilbrium steady state of the dissipative Bose-Hubbard model coupled to reservoirs on either end of the system by simulating the fixed point of the Lindblad master equation using the methods of time dependent DMRG, as well by providing analytical results for the noninteracting limits and for symmetric reservoirs. 
We have observed that in the case of one reservoir or more symmetric reservoirs, the system evolves
towards an uncorrelated product state, regardless of the interparticle interaction strength. We have confirmed the existence of decoherent-free subspaces when the system with an odd number of sites is coupled to a single reservoir. 
In the asymmetric setting, the density profile in the noniteracting case is flat in the bulk and evolves towards a linear profile with onset of interparticle interaction. 
We have also studied the driven-dissipative Bose Hubbard model subjected to coherent pumping and Lindblad reservoirs on either end of the system and we have shown that in the non-interacting case a sublattice of the system is protected from coherent pumping. The protection is gradually destroyed by increasing the interparticle interaction strength.

Our results suggest that the nonequilibrium steady state of the Bose-Hubbard model differs significantly from the equilibrium counterpart, especially in the case when a single site is coupled to a reservoir which after a long time destroys the overall correlations in the system. 
The existence of a subspace which is protected against coherent laser pumping on the other hand,  allows manipulation of sublattices. 

\begin{acknowledgments}
We thank M. Troyer, M. \v{Z}nidari\v{c}, and T. Prosen for discussions and comments on the manuscript as well as M. Dolfi, T. H. Seligman, I. Zintchenko, and J. Gukelberger for discussions.
This work was supported through the National Competence Center in Research (NCCR) QSIT. 
The simulations were run on the Brutus cluster at ETH Zurich.
\end{acknowledgments}

%%%%%%%%%%%%%%%%%%%%%%%%%%%%%%%%%%%%%%%%%%%%%
% Bibliography
%%%%%%%%%%%%%%%%%%%%%%%%%%%%%%%%%%%%%%%%%%%%%

\appendix
%%%%%%%%%%%%%%%%%%%%%%%%%%%%%%%%%%%%%%%%%%%%%
% APPENDIX A
%%%%%%%%%%%%%%%%%%%%%%%%%%%%%%%%%%%%%%%%%%%%%

\section{Exact solution for the noninteracting driven-dissipative Bose Hubbard model}
\label{app:A}
The noninteracting driven-dissipative Bose Hubbard model can be solved exactly in the framework of the ``third quantization'' \cite{prosen3q} and its recent extension to quadratic bosonic systems \cite{prosenseligman3q}. Here we present an explicit solution for the driven-dissipative bosonic model~(\ref{eq:H}) by following the formalism in Ref.~\cite{prosenseligman3q} and solving the relevant equations analytically. 

The Hamiltonian~(\ref{eq:H}) in the noninteracting limit $U=0$ can be written in a form
\[
H = -t \underline{a}^\dagger \cdot \mathbf{D} \underline{a} + \underline{\Omega} \cdot (\underline{a} + \underline{a}^\dagger )
\]
where $\mathbf{D}$ is a symmetric off-diagonal matrix with ones on the first off-diagonal, 
$D_{j,j+1} = D_{j+1,j} = 1$ and zeros elsewhere 
and $\underline{\Omega} \equiv \Omega (1,1,\ldots,1)$.
Following the formalism in Ref.~\cite{prosenseligman3q} or by setting $U=0$ in the general Liouvillian operator~(\ref{eq:liouvillian}) we obtain the Liouville operator in a form involving at most quadratic terms,
\begin{equation}
\hat{\mathcal{L}} = 
(\hat{\underline{a}},\hat{\underline{a}}') \cdot 
\begin{pmatrix} 
  \mathbf{0} & -\mathbf{X}^T \\
- \mathbf{X} & \mathbf{Y} \\
\end{pmatrix}
(\hat{\underline{a}},\hat{\underline{a}}') 
+ \underline{g} \cdot \hat{\underline{a}}'.
\label{eq:LXY}
\end{equation}
where matrices $\mathbf{X}$ and $\mathbf{Y}$ are parametrized by the pumping-free Hamiltonian and the reservoirs whereas coherent pumping contributes a linear term parametrized by 
the vector $\underline{g}$.
Comparing~(\ref{eq:LXY}) to~(\ref{eq:liouvillian}) we obtain 
\begin{eqnarray}
\mathbf{X} &=& -i \frac{t}{2} \sigma^3 \otimes \mathbf{D} + \frac{1}{2} \sigma^0 \otimes \textrm{diag}(\gamma_1 - \gamma_2) \nonumber \\
\mathbf{Y} &=& \sigma^1\otimes \textrm{diag}(\underline{\gamma_2}) \nonumber \\
\underline{g} &=& -i (\underline{\Omega},-\underline{\Omega})
\label{eq:XYg}
\end{eqnarray}
where $\sigma^{1,2,3}$ are the Pauli matrices and $\sigma^{0}$ is the identity matrix.
The main result of Ref.~\onlinecite{prosenseligman3q} for our purposes is that the two-point expectation values in the steady state are given as
\begin{equation}
\langle a_j^\dagger a_l \rangle = Z_{n+j, l} + s_{n+j} s_l
\label{eq:ajal}
\end{equation}
where $\mathbf{Z} \in \mathbf{C}^{2n \times 2n}$ is a solution of the \emph{Sylvester equation}
\begin{equation}
\mathbf{X}^T \mathbf{Z} + \mathbf{Z} \mathbf{X} = \mathbf{Y}
\label{eq:sylvester}
\end{equation}
and $\underline{s}$ is a solution of a linear problem
\begin{equation}
2 \mathbf{X}^T \underline{s} = \underline{g}.
\label{eq:Xs}
\end{equation}
This simple form of matrices $\mathbf{X}$ and $\mathbf{Y}$ allows us to obtain explicit solutions for the particle density profile in the steady state.

\subsection{Solution without coherent pumping}
First we consider the noninteracting bosonic model coupled to Lindblad reservoirs on either end of the system in the absense of the coherent pumping term $\Omega=0$. 
In this case the Liouvillian operator is a simple quadratic form with $\underline{g}=0$ and thus $\underline{s}=0$. It remains to solve the Sylvester equation~(\ref{eq:sylvester}). The block form of matrices
\begin{equation}
\mathbf{X} = \begin{pmatrix} \mathbf{\tilde X} & \mathbf{0} \\ \mathbf{0} & \mathbf{\tilde X}^* \end{pmatrix}
\quad
\textrm{and}
\quad
\mathbf{Y} = \begin{pmatrix} \mathbf{0} & \mathbf{\tilde Y} \\ \mathbf{\tilde Y} & \mathbf{0} \end{pmatrix} 
\label{eq:XYtilde}
\end{equation}
where 
$\mathbf{\tilde X} = \frac{1}{2}(\textrm{diag}(\underline{\gamma}_1-\underline{\gamma}_2)-i t \mathbf{D})$ 
and 
$\mathbf{\tilde Y} = \textrm{diag}(\underline{\gamma}_2)$
 allows us to express the complex symmetric matrix $\mathbf{Z}$ solving the Sylvester equation~(\ref{eq:sylvester}) in terms of a complex Hermitian matrix $\mathbf{\tilde Z}$ as 
\[
\mathbf{Z} = 
\begin{pmatrix} \mathbf{0} & \mathbf{\tilde Z} \\ \mathbf{\tilde Z}^T & \mathbf{0} \end{pmatrix} 
\]
by which the Sylvester equation is reduced to a \emph{continuous Lyapunov equation} 
\begin{equation}
\mathbf{\tilde X} \mathbf{\tilde Z} + \mathbf{\tilde Z} \mathbf{\tilde X}^H = \mathbf{\tilde Y}.
\label{eq:lyapunov}
\end{equation}

\subsubsection{Two boundary reservoirs}
The real parts of matrices $\mathbf{\tilde X}$ and $\mathbf{\tilde Y}$ can only have two non-zero elements (because there are only two reservoirs) which parametrize the bulk of the matrix $\mathbf{\tilde Z}$.  The set of $n$ linear equations~(\ref{eq:lyapunov}) thus reduces to a linear set of equations with four unknowns as
\[
\mathbf{\tilde Z}  =
\zeta_0^{-1} \begin{pmatrix}
z_L & i\zeta & 0 & \cdots  & 0 \\
-i \zeta & z_B & i \zeta & \ddots  & \vdots \\
0 & \ddots & \ddots  & \ddots  & 0 \\
\vdots & \ddots & -i \zeta & z_B & i \zeta \\
0 & \cdots & 0 &  -i \zeta & z_R
\end{pmatrix}
\]
The set of four equations for $\frac{z_L}{\zeta_0}, \frac{z_B}{\zeta_0}, \frac{z_R}{\zeta_0}, \frac{\zeta}{\zeta_0}$ can be solved analytically by means of e.g. Wolfram's Mathematica \cite{mathematica} and the solution can be written in a compact form by using the transformed parameters
$\kappa_j \equiv \gamma_{1,j}/\gamma_{2,j}-1$ and $\Gamma_j \equiv \gamma_{2,j} / t$ (see Eq.~\ref{eq:kappa}) as 
\begin{eqnarray}
z_L &=& 
(\Gamma_L^{-1} + \Gamma_R^{-1})  +  \kappa_R (\Gamma_L \kappa_L + \Gamma_R \kappa_R)
\nonumber \\
z_B &=& 
(\Gamma_L^{-1} + \Gamma_R^{-1})  +  (\Gamma_L \kappa_L^2 + \Gamma_R \kappa_R^2)
\nonumber \\
z_R &=& 
(\Gamma_L^{-1} + \Gamma_R^{-1})  + \kappa_L (\Gamma_L \kappa_L + \Gamma_R \kappa_R)
\nonumber \\
\zeta &=&
\kappa_L - \kappa_R
\nonumber \\
\zeta_0 &=& 
\frac{(\Gamma_L \kappa_L + \Gamma_R \kappa_R ) (\Gamma_L \Gamma_R \kappa_L \kappa_R + 1) }{\Gamma_L \Gamma_R}.
\end{eqnarray}
The expectation values~(\ref{eq:ajal}) are given by the lower-left block of the matrix $\mathbf{Z}$ 
and read (note that $\underline{s}=0$)
\[
\langle a_j^\dagger a_l\rangle = {\tilde Z}_{lj}
\]
which in turn results in the particle density 
\[
\langle n_j \rangle =
\begin{cases} 
z_L/\zeta_0 & \text{if $j=1$,} \\
z_B/\zeta_0 & \text{if $1<j<n$,} \\
z_R/\zeta_0 & \text{if $j=n$.}
\end{cases}
\]
The particle current, $J = i (a_{j+1}^\dagger a_{j} - a_j^\dagger a_{j+1})$ is associated with the expectation value of 
\[
\langle a_{j+1}^\dagger a_j \rangle = i \zeta / \zeta_0
\]
and vanishes in the equilibrium setting when $\kappa_L = \kappa_R$.

\subsubsection{One reservoir}
Let us now assume one reservoir coupled to the site $q$ in the system with Lindblad operators $L_1 = \sqrt{\gamma_1} a_q$ and $L_2 = \sqrt{\gamma_2} a_q^\dagger$. 
In this case the matrix $\mathbf{\tilde X}$ defined in~(\ref{eq:XYtilde}) can be written as 
\[
\mathbf{\tilde X} = \frac{\gamma_1 - \gamma_2}{2} \underline{e}_j \otimes \underline{e}_j - \frac{i t}{2} \mathbf{D}
\]
where $\underline{e}_q$ is a unit vector $(0,0,\ldots,0,1,0,\ldots,0)$ with a one at the position $q$.

The eigenvalues of the Liouvillian operator are determined through the eigenvalues of the matrix $\mathbf{X}$ (see Ref.~\cite{prosen3q} for details) which in turn are given by the eigenvalues of $\mathbf{\tilde X}$. The matrix $\mathbf{D}$ is a special tridiagonal Toeplitz matrix and its eigenvalues are given by $2 \cos(\frac{k \pi}{n+1})$ for $k=1,2,\ldots, n$ with the corresponding eigenvectors $[\underline{v}_k]_j = \sin(\frac{k j \pi}{n+1})$ for $j=1,2,\ldots,n$. 
It is easy to see that the kernel of $\mathbf{D}$ is non-empty if $n$ is odd and consists of a null eigenvector $\underline{w} = \underline{v}_{k=(n+1)/2}$ with elements $w_j = \sin(j \frac{\pi}{2})$. 
If the reservoir is attached to an even site $q$, then $\underline{w}$ is also the null vector of $\mathbf{\tilde X}$ and the kernel of $\mathbf{X}$ is at least of dimension two. This makes the null space of the Liouvillian operator infinitely dimensional and there exists not only one steady state but a manifold of steady states \cite{prosenspectral} forming a  protected decoherence free subspace. This subspace includes the product state given in Appendix~\ref{app:B}.
The decoherence free subspace thus exists for any odd system size with a reservoir attached to any even site in the system which does not necessarily include the site in the center of the system.

On the other hand, no decoherence free subspace exists for even system sizes $n$. 
This can be readily confirmed by 
calculating the determinant $\textrm{det}(\mathbf{\tilde X}) = (t/2)^{n}$ which is non-zero for any $q$ and any rates $(\gamma_1, \gamma_2)$. 
As a consequence, the steady state is a unique product state given in Appendix~\ref{app:B}.

The treatment presented here does not exclude the possibility of quasi-decoherence free subspaces which correspond to manifolds spanned by the eigenvectors of the liouvillian operator corresponding to completely imaginary eigenvalues (i.e. with zero real part). Such subspaces are not entirely insensitive to dissipation but accumulate a phase factor during the time evolution according to the master equation.
The treatment of steady state manifolds \cite{prosenspectral} is beyond the scope of this manuscript.

\subsection{Driven dissipative noninteracting}
The linear term in the Liouville operator~(\ref{eq:LXY}) can be removed by introducing new maps
\[
\underline{\hat{w}} = \underline{\hat{a}} - \underline{s} \hat{1}
\quad
\textrm{and}
\quad
\underline{\hat{w}}' = \underline{\hat{a}}'
\]
which again fulfil almost-canonical commutation relations $[\hat{w}_j,\hat{w}'_l] = \delta_{jl}$, $[\hat{w}_j, \hat{w}_l] = 0$ and $[\hat{w}'_j, \hat{w}'_l] = 0$.
The Liouville operator can now be rewritten in terms of $\underline{\hat{w}}$ and $\underline{\hat{w}}'$ as 
\[
\hat{\mathcal{L}} = - \underline{\hat{w}} \cdot \mathbf{X} \underline{\hat{w}}' - \underline{\hat{w}}' \cdot \mathbf{X}^T \underline{\hat{w}} +\underline{\hat{w}}' \cdot \mathbf{Y} \underline{\hat{w}}' +
(\underline{g}  - 2  \mathbf{X}^T \underline{s}) \cdot \underline{\hat{w}}'
\]
and the linear term vanishes by choosing the vector $\underline{s}$ as the solution of~(\ref{eq:Xs}).
The Liouville operator again takes a quadratic form~(\ref{eq:LXY}) with $\underline{g}=0$,
keeping the same matrices $\mathbf{X}$ and $\mathbf{Y}$ in~(\ref{eq:XYg}).
Due to a block structure~(\ref{eq:XYtilde}) of $\mathbf{X}$ and the vector $\underline{g} = -i (\underline{\Omega}, - \underline{\Omega})$, 
the linear equation~(\ref{eq:Xs}) can be reduced to a smaller set by using an ansatz $\underline{s} = (\underline{\tilde s}, \underline{\tilde s}^*)$ as 
\begin{equation}
2 i  \mathbf{\tilde X} \underline{\tilde s} = \underline{\Omega}
\label{eq:Xs2}
\end{equation}
where $\mathbf{\tilde X}$ was defined in~(\ref{eq:XYtilde}).
Again, due to the special form of $\mathbf{\tilde X}$ and $\underline{\Omega}$, the linear system of $n$ equations can be reduced to a linear system of at most $6$ equations for any system size. 
An explicit solution for $\underline{s}$ can then be used to obtain the particle density profile as given in~(\ref{eq:ajal}).

Let us for illustration solve the linear system~(\ref{eq:Xs2}) explicitly for the system size divisible by a factor of four. Let us assume that $\underline{s}$ is a real vector. Then 
$\textrm{Re}(\mathbf{\tilde X}) \underline{\tilde s}$ must vanish which implies ${\tilde s}_1 = {\tilde s}_n = 0$. 
On the other hand, $2 \textrm{Im}(\mathbf{\tilde X}) \underline{\tilde s} = - \underline{\Omega}$
implies that ${\tilde s}_{j-1} + {\tilde s}_{j+1} = \Omega/t$ which
immediately gives us ${\tilde s}_3 = \Omega/t$, ${\tilde s}_5 = 0$, ${\tilde s}_{7}=\Omega/t$ and so on, due to ${\tilde s}_1 = 0$. We repeat the same procedure starting from $n$ to $n-2, n-4, \ldots$ and 
obtain the solution 
\begin{equation}
\underline{\tilde s} = \frac{\Omega}{t} (0, 1,1, 0, 0, 1,1, 0,  \ldots, 0, 1,1,0).
\end{equation}
The correction to the particle density profile due to coherent pumping is  given by 
\[
\langle n_j \rangle = \langle n_j \rangle_{\Omega=0} + {\tilde s}_j^2
\]
and we immediately observe that only sites $2,3,6,7,10,11,\dots, n-2,n-1$ are affected by coherent pumping, yielding a correction of $(\Omega/t)^2$ to the particle density profile, 
while the other sites $1,4,5,8,9,\dots, n-4,n-3, n$ are protected from coherent pumping.

\begin{figure}
\centering
\includegraphics[width=\columnwidth]{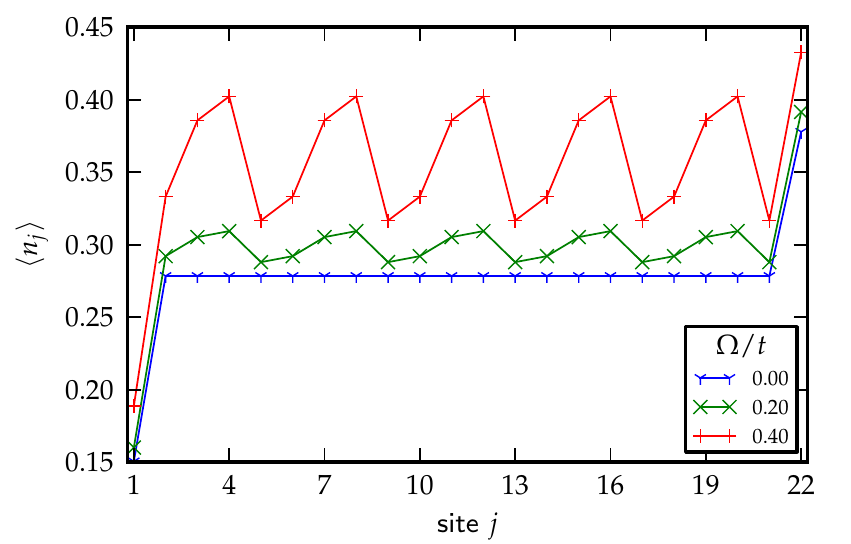}
\caption{(Color online) Particle density profile for a driven-dissipative noninteracting bosonic model~(\ref{eq:LXY}) on $n=22$ sites with 
$\kappa_L = 9$, $\kappa_R = 2.33$, $\Gamma_L = 0.2$, $\Gamma_R = 0.6$ and various $\Omega/t=0,0.2,0.4$ (bottom to top). 
}
\label{fig:Omega22}
\end{figure}

The generic case where the number of sites is not divisible by four can also be obtained explicitly by Mathematica \cite{mathematica}; the results however cannot be represented in a compact form. The crucial difference to the previously studied case is that the protection from coherent pumping disappears, except in the case of equal reservoirs on both edges (and an even number of sites). 
In Fig.~\ref{fig:Omega22} we show the particle density profile for $n=22$ sites and different reservoirs, $\underline{\gamma}_L = (1,0.1)$ and $\underline{\gamma}_R = (1,0.3)$ and $t=0.5$ where coherent pumping affects all sites; the zig-zag pattern however persists.

%%%%%%%%%%%%%%%%%%%%%%%%%%%%%%%%%%%%%%%%%%%%%
% APPENDIX B
%%%%%%%%%%%%%%%%%%%%%%%%%%%%%%%%%%%%%%%%%%%%%

\section{Exact solution for equal reservoirs}
\label{app:B}
The interacting term in the Liouvillian operator~(\ref{eq:liouvillian}) is a quartic term in operators
$(\underline{\hat{a}}, \underline{\hat{a}}')$ and the formalism of the third quantization no longer applies. In case of two or more equal reservoirs, 
however, the nonequilibrium steady state can still be obtained exactly as a product density matrix operator which we shall use as an ansatz,
\begin{equation}
\rho = \bigotimes_{j=1}^{n} {\tilde \rho_j}
\label{eq:gutzwiller}
\end{equation}
where ${\tilde \rho_j}$ is a single-site density operator,
\[
\xket{\tilde\rho_j} = 
\sum_{m_0, m_1=0}^{\infty} r_{m_0,m_1}\frac{  (\hat{a}_{0,j}')^{m_{0}} (\hat{a}_{1,j}')^{m_{1}} }{\sqrt{ m_{0}! m_{1}! } } \xket{\rho_0}.
\]
Such a Gutzwiller ansatz has proven useful in determining the equilibrium properties of the Bose Hubbard model and considering the fact that we assume an equilibrium setting with equal reservoirs, this makes a valid physical assumption. 

The Liouville operator consists of two parts: the unitary and the dissipative part. Here we shall assume that $\hat{\mathcal{L}} \ket{\rho}$ vanishes for both parts separately. First we consider the unitary part and observe that the contribution of the quartic term in~(\ref{eq:liouvillian}) involving an operator $\hat{n}_{0,j} - \hat{n}_{1,j}$ vanishes if $\hat{a}_{0,j}'$ and $\hat{a}_{1,j}'$ appear symmetrically in the steady state such that $r_{m_0,m_1} = \delta_{m_0,m_1} r_{m_0}$. 
The single site density operator ${\tilde \rho_j}$ thus reads
\[
\xket{\tilde\rho_j} = 
\sum_{m=0}^{\infty} r_{m} \xket{m,m}_j
\]
%with a normalization constraint $\sum_m r_m = 1$ 
where we have used an abbreviation 
\[
\xket{m_0,m_1}_j \equiv  \frac{(\hat{a}_{0,j}')^{m_0} (\hat{a}_{1,j}')^{m_1}}
{ \sqrt{m_0! m_1! } }  \xket{\rho_0}.
\]
Next we consider the tunneling term in the Liouvillian operator which maps a product operator for two neighboring $j$ and $j+1$  sites as
\begin{eqnarray}
 \xket{k,k}_j \otimes \xket{l,l}_{j+1} \mapsto
&-& t\sqrt{ (k+1)l} \xket{k+1,k}_j \otimes \xket{l-1,l}_{j+1} \nonumber \\
&+& t  \sqrt{ (k+1)l} \xket{k,k+1}_j \otimes \xket{l,l-1}_{j+1} \nonumber \\
&-& t \sqrt{ k (l+1)} \xket{k-1,k}_j \otimes \xket{l+1,l}_{j+1} \nonumber \\
&+& t \sqrt{ k (l+1)} \xket{k,k-1}_j \otimes \xket{l,l+1}_{j+1}. \nonumber
\end{eqnarray}
A straight forward calculation reveals that the tunneling part vanishes if 
$r_k /r_{k-1} = r_{l+1} / r_l$ for all $k$ and $l$. 
This condition is easily satisfied by choosing
\begin{equation}
r_k = \kappa^{-k} r_0
\label{eq:rk}
\end{equation}
for some positive real number $\kappa$. 
As mentioned earlier, the system amplifies indefinitely for $\kappa<1$  
whereas the steady state is trivial (equal to the maximally mixed state) for $\kappa=1$.
Finally, only the dissipative part of the Liouvillian operator~(\ref{eq:liouvillian}) remains and it vanishes if 
\[
r_{m} [ (\gamma_{j,1} - \gamma_{j,2}) 2 m  ] - 2 \gamma_{j,2} m r_{m-1} = 0
\]
for all sites $j$ that are coupled to reservoirs. It is easy to see that the recursive relation~(\ref{eq:rk})  indeed solves the above recursion relation by setting 
\[
\kappa = \gamma_{j,1}/\gamma_{j,2}-1.
\]
This finally gives us the requirements that have to be met in order that the product operator~(\ref{eq:gutzwiller}) is a solution for the steady state of the dissipative Bose Hubbard model.
It is required that $\kappa$ is site independent which is the case if (i) only one site is coupled to a reservoir, or (ii) the ratio $\gamma_{j,1}/\gamma_{j,2}$ is  constant for all reservoirs.

We can now immediately calculate the particle density in the steady state,
 given by ${\rm tr}[ a_j^\dagger a_j \rho]$ as 
\begin{equation}
\langle n_j \rangle = \frac{\bra{1} a_{0,j} a_{1,j} \ket{\rho}}{\braket{1}{\rho}} = 
\frac{r_1 r_0^{n-1}}{r_0^{n}} = \kappa^{-1}.
\end{equation}
This result is valid for any set of parameters in the Bose Hubbard model (including the disordered case with site-dependent tunneling rate and the interaction strength), if only the ratio between the emission rate and the absorption rate is constant for all existing reservoirs. Hence, this solution always applies to the case with a single reservoir

The result obtained in this section is valid for the Bose Hubbard model with an infinitely dimensional bosonic space. It is however also possible to obtain an equivalent result in the case of a truncated bosonic space where at most $K-1$ bosons are allowed per site. In such a case the nonequilibrium steady state is again given as~(\ref{eq:gutzwiller}) with 
\[
{\tilde \rho_j} = 
\frac{ \kappa (1 + \kappa)^{K-1}}{(1 + \kappa)^K-1}
\sum_{m=0}^{K-1} (\kappa+1)^{-m} \ket{m} {}_j {}_j\bra{m} 
\]
where 
\[
\ket{m}_j = \frac{(a_j^{\dagger})^m}{\sqrt{m!}} \ket{0}
\]
and the average number of particles obtains  a correction, exponentially falling with $K$, as 
\[
\langle n_j \rangle = \kappa^{-1} - \frac{K}{(1 + \kappa)^K-1}.
\]

This result offers a nice physical interpretation. First, two trivial limits, $\kappa=\infty$ (in case without particle injections) and $\kappa=0$ (without particle losses) correspond to a trivial vacuum state and the infinitely amplified state, respectively. In a nontrivial case, the reservoir imposes the desired particle density on the site it is coupled to whereas the hoppings make the system homogeneous. 
This phenomenon is surprising as it tells us that any initial state will eventually, albeit after a very long time if the tunneling rate is small, become completely uncorrelated, if the system is locally coupled to a reservoir and the steady state is unique. 
We note that the uniqueness is not guaranteed in general and there might exist other steady states in case of decoherence-free subspaces as shown in Appendix~\ref{app:A}.

\end{document}